\documentclass[aps,manuscript,superscriptaddress,nobibnotes]{revtex4-2}
\usepackage[utf8]{inputenc}
\usepackage{color}
\usepackage{amsmath}
\usepackage{graphicx}
\usepackage{subscript}
\usepackage[unicode=true,pdfusetitle,
bookmarks=true,bookmarksnumbered=true,bookmarksopen=true,bookmarksopenlevel=1,
 breaklinks=true,pdfborder={0 0 1},backref=false,colorlinks=true]{hyperref}
\usepackage{times}
\usepackage{xcolor}
\usepackage{hhline}

\setcitestyle{super}

\newcommand{\pej}[1]{{\color{black}#1}}

\begin{document}


\title{Giant photostriction in lead-free ferroelectric stemming from photo-excited thermalized carriers}

\author{Ga\"{e}lle Vitali-Derrien}
\affiliation{Universit\'{e} Paris-Saclay, CentraleSup\'{e}lec, CNRS, Laboratoire SPMS, 91190, Gif-sur-Yvette, France.}
\author{Oana Condurache}
\affiliation{Universit\'{e} Paris-Saclay, CentraleSup\'{e}lec, CNRS, Laboratoire SPMS, 91190, Gif-sur-Yvette, France.}
\author{Antoine Ducournau}
\affiliation{Universit\'{e} Paris-Saclay, CentraleSup\'{e}lec, CNRS, Laboratoire SPMS, 91190, Gif-sur-Yvette, France.}
\author{Pascale Gemeiner}
\affiliation{Universit\'{e} Paris-Saclay, CentraleSup\'{e}lec, CNRS, Laboratoire SPMS, 91190, Gif-sur-Yvette, France.}
\author{Maxime Vallet}
\affiliation{Universit\'{e} Paris-Saclay, CentraleSup\'{e}lec, CNRS, Laboratoire SPMS, 91190, Gif-sur-Yvette, France.}
\author{Nicolas Guiblin}
\affiliation{Universit\'{e} Paris-Saclay, CentraleSup\'{e}lec, CNRS, Laboratoire SPMS, 91190, Gif-sur-Yvette, France.}
\author{Thomas Antoni}
\affiliation{Universit\'{e} Paris-Saclay, ENS Paris-Saclay, CentraleSup\'{e}lec, CNRS, LuMIn, 91190, Gif-sur-Yvette, France.}
\author{Sylvia Matzen}
\affiliation{Universit\'{e} Paris-Saclay, CNRS, Centre de Nanosciences et de Nanotechnologies (C2N), 91120, Palaiseau, France.}
\author{Pascal Ruello}
\affiliation{Le Mans Universit\'{e}, CNRS, Institut des Mol\'{e}cules et Mat\'{e}riaux du Mans, 72085, Le Mans, France.}
\author{Dagmar Chvostova}
\affiliation{FZU- Institute of Physics, Czech Academy of Sciences, Na Slovance 2, 182 00 Praha 8, Czech Republic}
\author{Tetyana Ostapchuk}
\affiliation{FZU- Institute of Physics, Czech Academy of Sciences, Na Slovance 2, 182 00 Praha 8, Czech Republic}
\author{Jirka Hlinka}
\affiliation{FZU- Institute of Physics, Czech Academy of Sciences, Na Slovance 2, 182 00 Praha 8, Czech Republic}
\author{Simon Hurand}
\affiliation{Institut Pprime, Université de Poitiers, CNRS, ISAE-ENSMA, 86360, Chasseneuil-du-Poitou, France}
\author{Mouna Khiari} 
\affiliation{Laboratoire de Physique de la Mati\`{e}re Condens\'{e}e, Universit\'{e} de Picardie Jules Verne, 80039, Amiens, France.}
\author{Houssny Bouyanfif}
\affiliation{Laboratoire de Physique de la Mati\`{e}re Condens\'{e}e, Universit\'{e} de Picardie Jules Verne, 80039, Amiens, France.}
\author{Charles Paillard}
\email{paillard@uark.edu}
\affiliation{Smart Ferroic Materials Center, Institute for Nanoscience \& Engineering and Department of Physics, University of Arkansas, Fayetteville AR 72701, USA.}
\author{Pierre-Eymeric Janolin}
\affiliation{Universit\'{e} Paris-Saclay, CentraleSup\'{e}lec, CNRS, Laboratoire SPMS, 91190, Gif-sur-Yvette, France.}

\begin{abstract}
\textbf{Ferroelectrics are polar materials whose polarization can be switched by applying electric fields; they offer unique opportunities to develop performant photostrictive materials, \textit{i.e.}, materials \pej{that} 
can deform under visible light illumination. 
Naturally devoid of inversion symmetry, they exhibit original photogalvanic effects such as the Bulk Photovoltaic Effect, which relies on ``hot'' photoexcited carriers. 
It has long been thought that the electric field generated by this effect may couple to the natural piezoelectric abilities of ferroelectrics to provide large photoinduced deformations. 
However, due to competing effects, such as thermal dilatation, deformation potential, polarization, or depolarizing-field screening by \textit{thermalized} carriers, it remains unclear \pej{which} 
microscopic phenomena govern the photoinduced deformations in classical ferroelectric materials. 
Here, we demonstrate the largest photoinduced deformation measured in a ferroelectric thin film. Reaching 1\%, this giant photostriction likely originates from the contribution of thermalized photoinduced carriers.}
\end{abstract}
\maketitle

\section*{Introduction}

\begin{figure}
    \centering
    \includegraphics[width=0.8\linewidth]{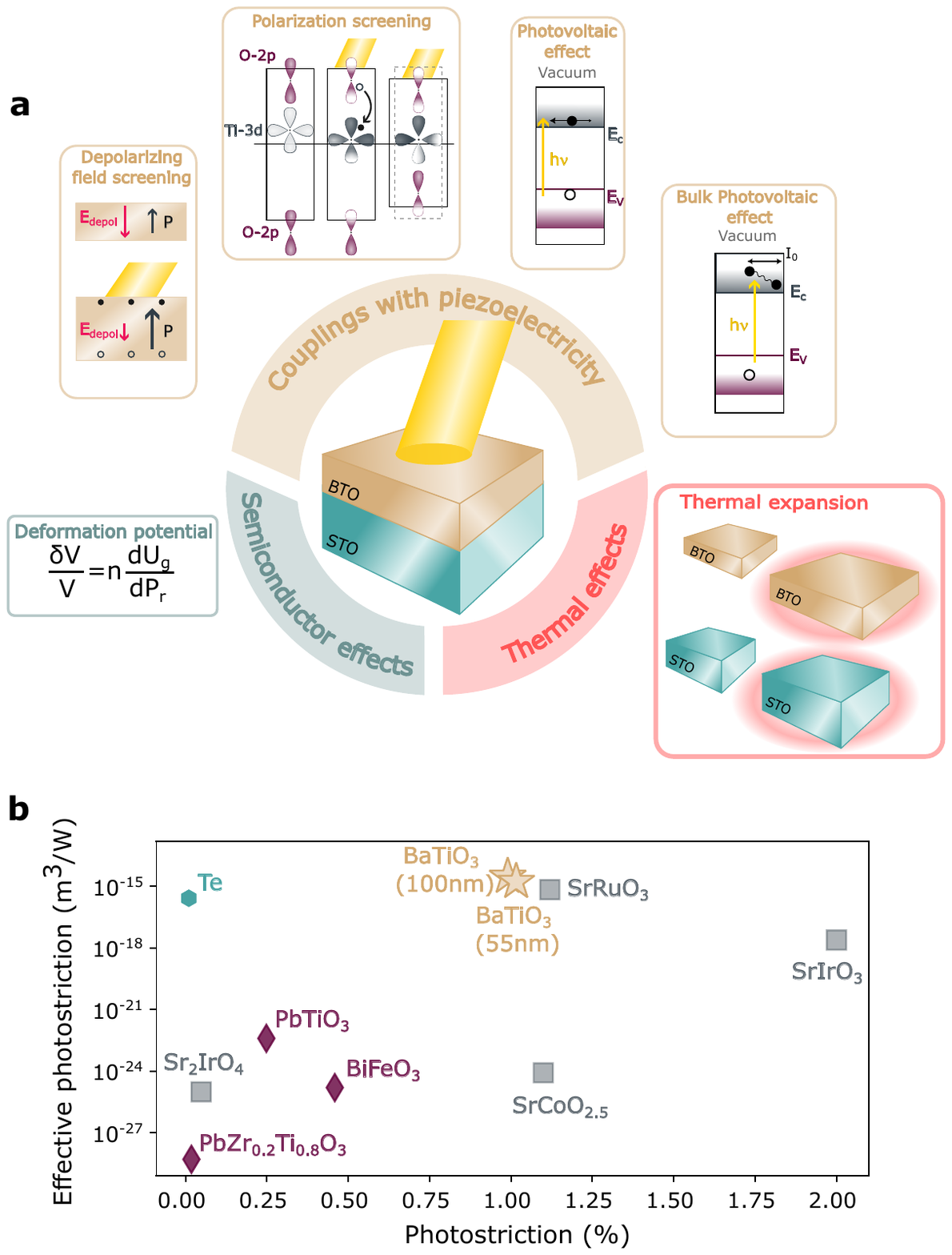}
    \caption{\textbf{Giant photostriction in barium titanate thin films.} \textbf{(a)} Microscopic mechanisms involved in the quasistationary photostrictive response. \textbf{(b)} Figure of Merit of the photostrictive response in sub-micrometer inorganic thin films, showing the competitive edge of (001)-BaTiO\textsubscript{3} films grown on SrTiO\textsubscript{3} (beige stars). Ferroelectric materials are represented as purple diamonds, while non-ferroelectric perovskites are in grey squares. Trigonal tellurium is represented as a blue hexagon. Note that the timescales may vary from ultrafast to quasistationary.}
    \label{fig:1}
\end{figure}

Ferroelectrics form a distinct class of functional materials due to their inherent switchable electrical polarization. 
This particular feature leads to strong electron-lattice couplings as manifested by their piezoelectric nature, the generation of ultrafast, strong, coherent acoustic waves~\cite{Lejman2014,Gu2023}, and large mechanical deformations~\cite{Daranciang2012,Wen2013,Schick2014} under visible or near-visible optical excitation. Therefore, ferroelectrics have been deemed extremely promising to generate large photoinduced mechanical deformations, a phenomenon known as photostriction. Potential applications cover optically-driven micro- or nanorobots~\cite{Poosanas2000}, optical switches and actuators~\cite{Ganguly2024}, or optomagnetic memories~\cite{Iurchuk2016}.


Several studies have explored the photostriction of ferroelectrics, starting with PbTiO\textsubscript{3}~\cite{Daranciang2012}, and PbTiO\textsubscript{3}-based solid solutions~\cite{Poosanaas1999} that gathered interest due to a peak in their piezoelectric response at the morphotropic phase boundary. 
Photostriction of multiferroic BiFeO\textsubscript{3} has also been thoroughly investigated at ultrafast~\cite{Lejman2014,Gu2023,Wen2013,Schick2014} and quasistatic~\cite{Kundys2010} time scales. 
The photoinduced strain in such inorganic ferroelectric materials is typically of the order of 0.1\%; meanwhile, their effective photostriction (defined as the total displacement divided by the incident light surface power), which translates \pej{to} the efficiency of the conversion of light into a mechanical displacement, is remarkably low (below 10\textsuperscript{-20}\,\pej{m$^3$/W}). 
Meanwhile, other \pej{(non-ferroelectric)} inorganic materials have already broken the 1\,\% photoinduced strain mark, with efficiencies as high as 10\textsuperscript{-15}\,\pej{m$^{3}$/W} (see Figure~\ref{fig:1}b)\pej{, far surpassing ferroelectrics so far.}

Surprisingly, photostriction in BaTiO\textsubscript{3} (BTO), the 
prototypical ferroelectric material, has been far less explored, except for a few recent works on free-standing membranes~\cite{Ganguly2024}, transient response~\cite{Hoang2025}, or first-principle calculations~\cite{Paillard2017}. 
In contrast to many inorganic \pej{ferroelectrics exhibiting} 
significant photostrictive response, such as lead- or bismuth-based compounds~\cite{Poosanaas1999,Kundys2010,Zhou2016}, barium titanate is non-toxic and environmentally friendly.
Thus, there is an urgent need to assess the \pej{specific} potential of barium titanate for photostrictive applications and to better understand the dominant mechanism \pej{underlying} 
photostriction in ferroelectrics. 
In this work, we demonstrate a giant photostriction in barium titanate thin films reaching 1\%, and discuss its potential origin.

Photostriction in ferroelectrics involves multiple, concomitant microscopic mechanisms that are depicted in Figure~\ref{fig:1}a. 
Some of them are not related to the ferroelectric polarization, such as thermal expansion (which is due to the photogenerated heat), and deformation potential (which is due to the semiconductor nature of the material, and involves the bandgap pressure sensitivity~\cite{Kundys2015}). 
Ferroelectricity \pej{offers} 
additional 
conversion \pej{mechanisms} of absorbed optical photons because of the natural asymmetry endowed by their spontaneous electric polarization and its strong coupling with mechanical degrees of freedom. 
For instance, \pej{our} first-principles calculations have predicted that thermalized photo-excited carriers could screen the electric polarization, resulting in a piezoelectric deformation of the lattice~\cite{Paillard2016,Paillard2017}.
In thin films, experimental and theoretical reports indicate that these photo-excited carriers in quasi-equilibrium mitigate \pej{depolarizing-field effects} 
by migrating to the interfaces, thereby inducing a piezoelectric-related deformation~\cite{Matzen2019,Dansou2022} \pej{characterized by the $d$ coefficient}. 
Additionally, the hot, non-thermalized portion of photo-excited carriers can generate strong photovoltages in excess of the bandgap~\cite{Glass1974} by means of the anomalous Bulk Photovoltaic Effect (BPVE) in noncentrosymmetric materials~\cite{Sturman1992} \pej{such as} 
ferroelectrics. 
Such large photovoltages (and those created by more conventional photovoltaic effect related to Schottky contacts~\cite{Paillard2016-review}) can also couple with the piezoelectric effect \pej{($d$ constant)} to induce a photoinduced mechanical deformation. 
Thus, it is 
unclear whether \textit{(i)} thermalized or non-thermalized photo-excited carriers and \textit{(ii)} interfacial or bulk-like effects dominate over the photoinduced mechanical response of \pej{ferroelectrics}. 
Here, we demonstrate a photoinduced strain exceeding 1~\% in a BaTiO\textsubscript{3} \pej{thin film} and an effective photostriction of 3$\times$10$^{-15}$\,m$^3$/W, constituting the largest photostrain in ferroelectrics, and the largest effective photostriction in perovskite oxide and semiconductor thin films (see Figure~\ref{fig:1}b). 
Furthermore, our analysis \pej{reveals that the photostrictive} 
strain measured in BaTiO\textsubscript{3} does not originate from thermal or photovoltaic effects, but 
from thermalized photo-excited carriers screening the polarization.



\section*{Results}

\begin{figure}[h]
    \centering
    \includegraphics[width=\linewidth]{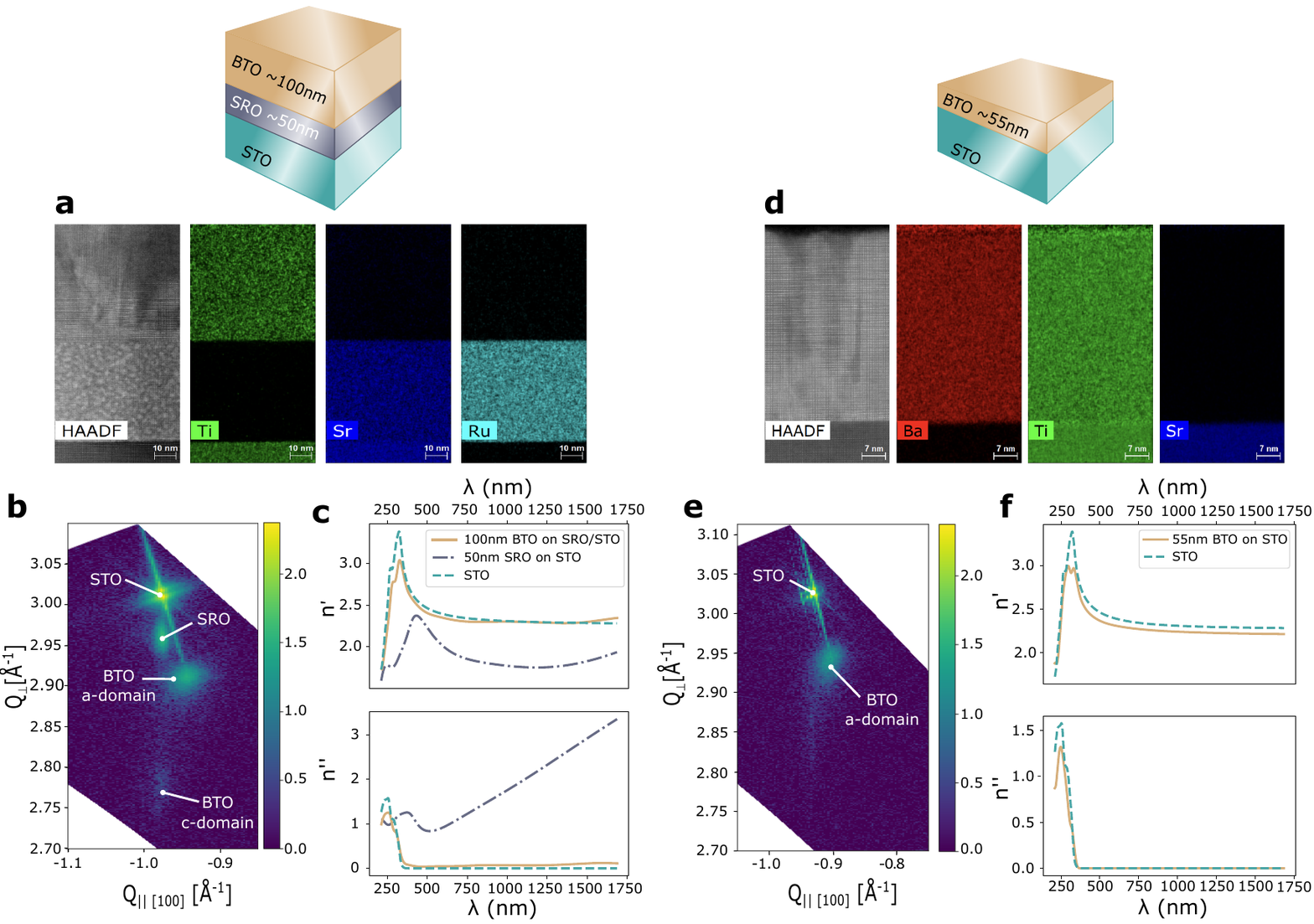}
    \caption{\textbf{Barium titanate films structures and optical properties.} \textbf{(a, d)} High-resolution transmission electron microscopy and electron X-ray dispersive microscopy of a 100~nm BaTiO\textsubscript{3} film on a 50~nm SrRuO\textsubscript{3} buffer layer and of a 55~nm BaTiO\textsubscript{3} film \pej{showing clear interfaces}; \textbf{(b, e)} Reciprocal space maps of the (103) plane for the 100~nm BaTiO\textsubscript{3} on 50~nm SrRuO\textsubscript{3} and 55~nm BaTiO\textsubscript{3} films \pej{exhibiting the coherent growth and domain structure}; \textbf{(c, f)} Optical index of the 100~nm BaTiO\textsubscript{3}, the 50~nm SrRuO\textsubscript{3}, the 55~nm BaTiO\textsubscript{3} and the SrTiO\textsubscript{3} substrate obtained by ellipsometry.}
    \label{fig:carac}
\end{figure}

To uncover the underlying mechanisms of photostriction in barium titanate, we investigate the photoinduced mechanical deformation of a 100~nm 
BaTiO\textsubscript{3} thin film, epitaxially grown by Pulsed Laser Deposition on top of a (001)-oriented SrTiO\textsubscript{3} (STO) substrate with a 50~nm thick SrRuO\textsubscript{3} (SRO) \pej{bottom electrode}
, and of a 
55~nm 
(001)-oriented BTO film directly grown on a STO substrate (see Methods). 
The SRO electrode enables \pej{the measurement of} 
electrical properties but \pej{complicates} 
the 
analysis. 
Therefore, 
only 
one of the two samples \pej{uses it}. 

Structurally, high-resolution scanning transmission electron microscopy images show columnar growth of the BaTiO\textsubscript{3} films in Figure~\ref{fig:carac}a and d, with little to no interface intermixing based on Electron X-ray Dispersive Spectroscopy. 
\pej{The relative intensities of the (103) peaks shown on the}
X-ray diffraction reciprocal space maps 
in Figure~\ref{fig:carac}b and e indicate 
\pej{that the 100-nm BTO films consists of} 
about 98~\% \textit{a}-domains and 2~\% \textit{c}-domains 
\pej{whereas} the thinner 55~nm film 
(directly deposited onto SrTiO\textsubscript{3}) only displays \textit{a}-domains.

The real and imaginary parts of the refractive index were determined by ellipsometry on both samples, for wavelengths ranging from 200~nm to 1,700~nm, in Figures \ref{fig:carac}c and f. 
As expected for BTO and STO, the extinction coefficient is low in the visible range, but can be increased due to defects (which is especially the case in the thicker BTO film, as can be seen from the growth in Figure~\ref{fig:carac}a).


The photoresponse of our films \pej{was investigated under illumination} 
with a 405-nm continuous wave ``pump'' laser whose intensity is modulated sinusoidally at a frequency of 10~Hz. 
The \pej{resulting} photoinduced displacement \pej{was measured} with sub-nanometer sensitivity by interferometry,  employing a second \pej{(``}probe\pej{''}) laser operating with a 594-nm wavelength (see Methods). 
Photoinduced displacements of a 50-nm SRO \pej{electrode} 
grown on STO and of a \pej{bare STO} substrate 
were also quantified to serve as references and, later on, \pej{to} isolate the individual response of the BTO \pej{films}.

\begin{figure}
    \centering
    \includegraphics[width=\linewidth]{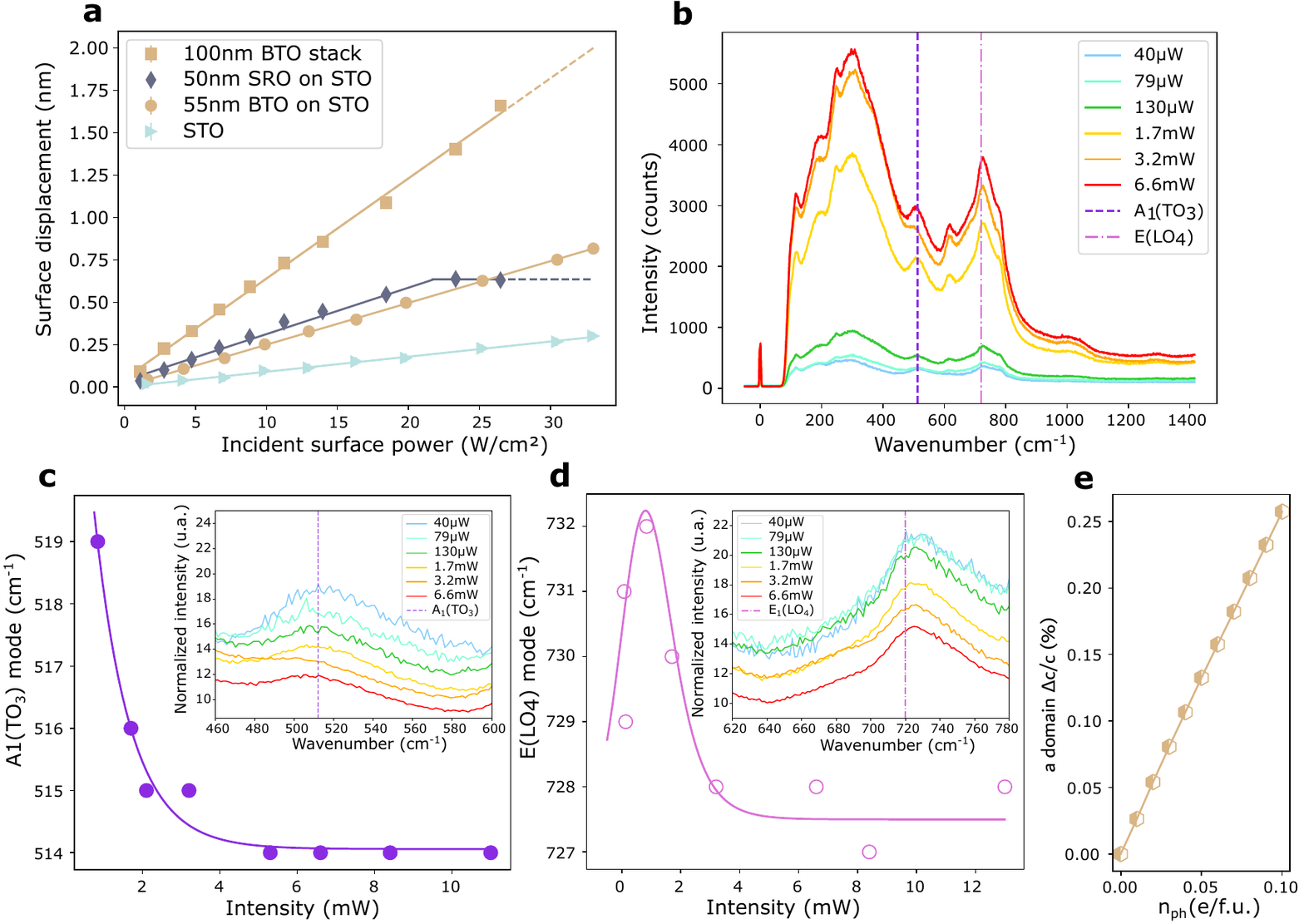}
    \caption{\textbf{Photostrictive response of barium titanate films.} \textbf{(a)} Photoinduced surface displacements of the 100-nm and 55-nm thick barium titanate (BTO) thin films, 50-nm strontium ruthenate (SRO) and bare strontium titanate (STO) substrate under increasing laser intensity at 405~nm. One set of measurements (STO, BTO/STO) ranges from \pej{1.6\,W.cm$^{-2}$ to 33\,W.cm$^{-2}$}, and the other (SRO/STO, BTO/SRO/STO) goes from \pej{1.1\,W.cm$^{-2}$ to 26.5\,W.cm$^{-2}$}, and the expected responses up to 33\,W.cm$^{-2}$ are in dotted lines. \textbf{(b)} Raman spectra of the 100-nm BTO/SRO/STO sample under varying power. \textbf{(c, d)} Variation of the A\textsubscript{1}(TO\textsubscript{3}) (resp. E(LO\textsubscript{4})) mode position as a function of increasing power, with associated fits; the inset shows a zoom on the normalized spectra. \textbf{(e)} photoinduced change in the out-of-plane lattice constant versus photo-excited carrier concentration $n_{ph}$ calculated by Density Functional Theory in $a$-domain tetragonal BTO.}
    \label{fig:3}
\end{figure}

The photoinduced displacement of the 100-nm BTO/SRO/STO sample increases linearly with 
the amplitude of the sinusoidally modulated intensity of the pump laser 
(see Figure~\ref{fig:3}a). 
The maximum 
surface displacement is 1.66~nm \pej{at} 26.5~W.cm\textsuperscript{-2} 
and is 
limited \pej{only} by the pump laser\pej{'s maximum power} as 
\pej{no} saturation 
\pej{is observed}. 
In contrast, the photoinduced surface displacement of the SRO/STO system (\textit{i.e.}, without \pej{the BTO} 
film) reaches a maximum photoinduced surface displacement 
\pej{of only} 0.63~nm \pej{(\textit{i.e.}, $\approx$38\,\% of the BTO/SRO/STO system) and saturates, while } 
the bare STO substrate \pej{exhibits an even smaller photoinduced displacement of} 0.23~nm \pej{($\approx$14\,\% of the BTO/SRO/STO system)} at a comparable laser fluence. 
Therefore, \pej{each component of the BTO/SRO/STO sample contributes to the measured photoinduced displacement, including the conducting SRO~\cite{Wei2017}. Still,}
a significant portion of the photostrictive response actually originates from the polar barium titanate layer.

\pej{In comparison, t}
he surface displacement of the 55~nm BTO film \pej{deposited directly onto STO} reaches 0.82~nm when the amplitude of the 405~nm pump surface intensity is 33~W.cm\textsuperscript{-2} (see Figure~\ref{fig:3}a), while the bare STO substrate's surface \pej{displacement is }
0.30~nm. \pej{This smaller photoinduced displacement of the BTO/STO sample (about half that of the BTO/SRO/STO sample) can be attributed to the absence of the SRO electrode (removing its contribution and its heating of both the BTO film and the STO susbstrate), the absence of minority $c$-domains (see Figure~\ref{fig:carac})a}, and the reduced thickness.

Note that the results in Figure~\ref{fig:3} cannot be directly converted into a photoinduced strain. 
Indeed, both the substrate and the SRO electrode (when present) contribute to the total photoinduced surface displacement. 
\pej{What is more, the photoinduced displacement }
encompasses the photostrictive (\textit{i.e.}, athermal) deformation 
and \pej{the} thermal dilatation of each layer (as shown in Figure~\ref{fig:3}a). 
\pej{Therefore, the power transmitted through the BTO film into the SRO electrode (when present) and the STO substrate must be appropriately modelled to separate the contributions of each layer to the total photoinduced displacement.} 

\pej{The power distribution in the samples layers depends on the optical properties of each component, which have been measured by ellipsometry } 
(see Figure~\ref{fig:carac}c-f). 
The refractive index at the pump wavelength (\textit{i.e.}, 405~nm, $\hbar\omega$=3.06~eV) of BTO ($n^{\prime}_{\rm BTO}$) is $\approx 2.5$ for both BTO films and the extinction coefficients ($n^{\prime\prime}_{\rm BTO}$) are around 0.06 and 0.01 for the 100-nm and 55-nm films, respectively. 
Hence, in the case of the 55-nm BTO/STO sample, 70~\% of the power that reaches the STO surface in the air/STO/air configuration penetrates 
the substrate in the air/BTO/STO/air case (see Supplementary Information). 
At the maximum investigated surface power 
\pej{(}33 W.cm\textsuperscript{-2}\pej{)}, only 23.1~W.cm\textsuperscript{-2} penetrate the substrate 
\pej{through} the BTO/STO interface. 
Based on the linear \pej{photoinduced displacement response of the bare} substrate 
(see blue curve in Figure~\ref{fig:3}a), 
\pej{these 23.1~W.cm\textsuperscript{-2} generate a photoinduced displacement of} $d^{\rm illumination}_{\rm STO}$=0.21~nm, \pej{\textit{i.e.}, about a quarter of the total photoinduced displacement of the BTO/STO system.}
\pej{For the BTO/SRO/STO sample, t}
he same model shows that more than 95~\% of the incident light reaching the SRO electrode in the air/SRO/STO/air configuration reaches the SRO electrode below the 100-nm BTO film. 
\pej{The higher transmission, despite the larger thickness, is due to different interfacial conditions.} 
\pej{From Figure~\ref{fig:3}a (grey curve), the corresponding photodinduced displacement of the SRO/STO sample is } 
$d^{\rm illumination}_{\rm SRO/STO}=$0.62~nm\pej{, \textit{i.e.}, about 38\% of} 
the total surface displacement of 1.66~nm of the BTO/SRO/STO sample.  
\pej{Therefore, the photoinduced displacement of the BTO films, for both thicknesses, contributes predominantly to the total photoinduced surface displacement.}

\pej{However, such displacements are both thermal and athermal, as part of the absorbed power is transformed into heat within each layer, contributing to the total surface displacement }
(see Figure S1 step 3)\pej{.} 
\pej{T}he temperature increase at the surface of the BTO film (at the pump laser modulation frequency) 
\pej{has been measured with an infrared camera to be }
6~mK (the detailed estimation is given in the Supplementary Material) for both BTO/STO and BTO/SRO/STO samples. \pej{Such a temperature increase corresponds to a thermal contribution of the BTO film to the total displacement of $d^{\rm temperature}_{\rm BTO}$=0.002\,nm (2\,fm).}
\pej{As the}
characteristic temperature diffusion time in the BTO films 
\pej{(}of the order of 10~ns, 
see Supplementary Materials)\pej{, which is much shorter than the}  
typical time (0.1s) of modulation of the pump laser intensity, \pej{the temperature is assumed to be uniform in the BTO film, leading to }
a thermally-induced surface displacement of $d^{\rm temperature}_{\rm STO} = d^{\rm temperature}_{\rm SRO/STO} = 0.05 $~nm in both systems under a 33~W.cm\textsuperscript{-2} incident surface power. 


\pej{T}o determine the sign of the deformation, \pej{w}
e performed power-dependent Raman spectroscopy on the 
100-nm film 
(see Figure \ref{fig:3}b). \pej{Consistent with the reciprocal maps, both the }
A\textsubscript{1}(TO\textsubscript{3}) mode at 515~cm\textsuperscript{-1}, due to the presence of \textit{a}-domains \cite{Rubio-Marcos2015} and the E(LO\textsubscript{4}) mode \pej{at 730~cm\textsuperscript{-1}}
\pej{related to \textit{c}-domains} are visible. 
The redshift of both modes under increasing laser power at 405~nm (as shown in Figures \ref{fig:3}c-d) indicates an overall dilatation under illumination (see Ref~\cite{Aldo2015} and complementary Density Functional Theory calculations in the Supplementary Information). 
An increase of the out-of-plane lattice parameter of our \pej{(mostly, for the 100-nm or fully, for the 50-nm)} \textit{a}-domain BTO films is thus expected under illumination, contributing to the measured photoinduced displacements of the surface in both BTO systems (Figure \ref{fig:3}a). 
Positive surface displacement \pej{(extension)} is also expected in SRO~\cite{Wei2017} and in STO due to its positive bandgap pressure sensitivity \cite{Khaber2014} and thermal expansion.

\pej{Therefore, t}
he surface displacement contributions emerging from all the layers have the same sign: photostriction of BTO, thermal and athermal response from the substrate (and \pej{SRO electrode} 
when present) lead to an overall expansion. 
\pej{S}ince the surface displacement of both the BTO film and the STO substrate have the same sign, the effective surface displacement of the 55~nm BTO film alone is $ d_{\rm BTO-55\,nm} = d^{\rm total}_{\rm BTO/STO} - d^{\rm illumination}_{\rm STO} - d^{\rm temperature}_{\rm STO} =$ 0.82 - 0.21 - 0.05 = 0.56~nm at the maximum investigated surface power of 33~W.cm\textsuperscript{-2}. 
Accordingly, the light induces a giant strain in the \pej{sole} 55-nm BTO film 
of 1.01~\%. 
In the more complex case of the 100-nm 
BTO/SRO/STO stack, a similar analysis leads to an estimated photoinduced surface displacement of the \pej{sole} BTO film 
of 0.99~nm under 26.5~W.cm\textsuperscript{-2}, 
corresponding to a giant photostrictive strain of 0.99~\%. 
Such large strains set a new record in ferroelectric films, surpassing the \pej{previously reported} 
0.6~\% photoinduced strain in BiFeO\textsubscript{3} \pej{measured by }
ultrafast pump-probe experiments~\cite{Schick2014}. 
Similarly, the effective photostriction \pej{of about }
$3\times 10^{-15}$~m\textsuperscript{3}/W exceeds most reported values in inorganic films (see Figure~\ref{fig:1}b). 





\newpage 

\section*{Discussion}


\pej{Having isolated the contributions from the SRO electrode (for the 100-nm film) and from the STO substrate, }
we discuss \pej{hereafter the} 
relative contributions \pej{of the various phenomena (thermal expansion, photovoltaic effects, couplings with piezoelectricity, or thermalized photoexcited carriers effects, 
\pej{illustrated in} Figure \ref{fig:1}) that may be at the origin of} 
the photo-generated deformation \pej{in the BTO film alone}

As \pej{for SRO and STO}, 
an obvious contribution \pej{to the deformation under optical excitation} is the thermal expansion \pej{of the BTO film. }
\pej{In previous reports~\cite{Kundys2010, Kundys2011} t}
his contribution has 
been ruled out due to dynamical response considerations
, \textit{i.e.}, the thermal expansion is 
expected to be much slower than the observed photostriction. 
However, as mentioned previously, the characteristic thermal response time $\tau_{th} \approx 10$~ns in our BTO thin films 
is much shorter than the modulation time of the pump laser (typically 0.1~s). 
From the \pej{measured surface} temperature increase $\Delta T_{\rm BTO}(z=0) = $ 6~mK under 33~W.cm\textsuperscript{-2}, we modelled the temperature profile $\Delta T_{\rm BTO}(z)$ in the 55~nm BTO thick film (see Supplementary Information). 
\pej{Consistent with our hypothesis, the temperature in the BTO film can be considered }uniform as the temperature 
\pej{difference throughout the film is only }
0.002~mK (\textit{i.e.}, 0.03~\%). 
\pej{T}he \pej{corresponding} thermal expansion displacement contributing to the photoinduced displacement of the BTO film amounts to $d_{\rm BTO-55~nm}^{\rm temperature}=\int_0^{55\,\rm nm} \alpha_{\rm BTO}\, \Delta T_{\rm BTO}(z)\, {\rm d}z$\,=\,2$\cdot$10$^{-6}$\,nm 
(2~fm). 
This is negligible compared to the total photoinduced displacement measured in the BTO film \pej{(0.56\,nm)}. 
Similarly, in the 100-nm BTO film, we estimate that the thermal expansion to $d_{\rm BTO-100~nm}^{\rm temperature}$\,=\,4$\cdot$10$^{-6}$\,nm 
(4~fm). 
This, too, is negligible compared to the 
\pej{measured} displacement \pej{(0.99\,nm)}. 
The thermal expansion of the BTO films is\pej{, therefore,} a minor contribution to the photostrictive response of BTO.

Photovoltaic effects, such as those generated by the Bulk Photovoltaic Effect (BPVE)~\cite{Sturman1992}, generate electric fields strong enough \pej{that} they would induce a sizable mechanical deformation through the reverse piezoelectric effect \pej{and, therefore, would contribute to the photoinduced surface displacement.}
To estimate this contribution, we performed transport measurements under illumination, and \pej{we} investigated the piezoelectric response (see Methods and Supplementary Information) on the 
100-nm BTO film \pej{(the one with} 
a bottom SRO electrode\pej{)}. 
The obtained open-circuit photovoltage saturates with light intensity at about 1~V. 
Such photovoltage is of the same order of magnitude as the giant photovoltaic response reported in thinner barium titanate films by Zenkevich \textit{et al.}~\cite{Zenkevich2014}. 
\pej{The direct piezoelectric response was measured from the interferometric measurement of the }
electric-field-induced surface displacement. 
\pej{The measured} effective out-of-plane longitudinal piezoelectric constant $d^*$\,$\approx$\,5~pm/V on the 100-nm thick BTO sample (see Methods and Supplementary Information). 
\pej{Therefore, }
the photoinduced displacement due to photovoltaic effects, $d_{\rm BTO}^{\rm PV}$, 
amounts to $ \approx 5 \text{~pm/V} \times 1 \text{~V} \approx\,$\pej{0.005\,nm (5\,pm)}. 
Even though the presence of 50~x~50~µm\textsuperscript{2} metallic pad serving as a top electrode changes the boundary conditions compared to the photostriction measurement, 
the displacement due to photovoltaic effects is orders of magnitude smaller than the measured total photoinduced displacement of 0.56~nm in the 55-nm BTO film and 0.99~nm in the 100-nm BTO film. Therefore, photostriction arising from a \textit{macroscopic} photovoltaic effect is unlikely to result in significant deformations. 

\pej{I}t could be argued that \textit{microscopic} \pej{(\textit{i.e.},} domain-confined\pej{)} photovoltaic and piezoelectric effects lead to significant macroscopic deformations in piezoelectric PMN-PT~\cite{Liew2022}. 
In our samples, this mechanism would mean that absorbed photons create ``hot'' photo-excited carriers which, by virtue of the BPVE, create a local open-circuit photovoltage $V_{\rm oc}^{\rm domains}$ inside the $a$-domains, whose characteristic lateral size is expected to be of the order of 
1.2~µm~\cite{Yuan2013}. 
From theoretically reported values of the BPVE shift current tensor~\cite{Young2012, Fei2020, Dai2021}, the photovoltage $V_{\rm oc}^{\rm domains}$ is estimated to reach $\sim 5$~mV. 
This would produce a photoinduced strain of the $a$-domains of $\sim 1.4 \times 10^{-7}$, \textit{i.e.}, orders of magnitude smaller than \pej{the 1\% strain} 
we observe (see Supplementary Materials). 
In contrast to various reports, our analysis indicates that, in the present case, the contribution from photovoltaic effects, 
\pej{whether} macro- or micro-scopic, conventional or related to ``hot'' (non-thermalized) photo-excited carriers, is negligibly small.

The photostrictive response thus likely stems from the actions of \textit{thermalized photo-excited carriers}
, which include the polarization screening~\cite{Khomskii2016} and deformation potential~\cite{Gauster1969} mechanisms. 
The polarization screening mechanism is illustrated in Figure~\ref{fig:1}: in barium titanate, the polarization 
\pej{stems} from the hybridization of the O-2$p$ and Ti-3$d$ orbitals. These orbitals are respectively involved in the valence and conduction bands so that, under illumination, an electron is transferred from the oxygen to the titanium atom. 
\pej{This decreases the off-centering of the titanium ion} and \pej{induces} an overall modification \pej{of the }structure \pej{through the piezoelectric effect characterized by the $g$ constant.} 
\pej{The second action of thermalized photo-carriers is t}
he deformation potential mechanism \pej{that} relies on the bandgap pressure sensitivity \pej{of any semiconductor}, triggering an anisotropic volume change when carriers are photogenerated.

To ascertain the effects of thermalized photoexcited carriers, we performed Density Functional Theory calculations of BTO $a$-domains clamped to the calculated lattice constant of STO, meanwhile enforcing a concentration $n_{ph}$ of electrons (respectively, holes) in the conduction (respectively, valence bands)
~\cite{Paillard2019, Abinit2025}. 
Our results (see Figure~\ref{fig:3}e) indicate that the out-of-plane lattice constant of $a$-domains increases with a rate of approximately 1.5\,$\times$\,10$^{-24}$~cm\textsuperscript{3}/number of electron-hole pairs.
\pej{To reach the non-thermal displacement of 0.56~nm measured in the 55-nm BTO film, the} 
concentration of photo-excited thermalized carriers \pej{would need to be} 
6.2\,$\times$\,10$^{21}$~electron-hole pair/cm\textsuperscript{3}. 
Based on a simple two-level model, such a concentration of photo-excited carriers is realistic with a recombination time of about 2~ms ~\cite{Magnan2020} (see Supplementary Information). 
This 
high recombination time \pej{implies} 
that our film contains defects, which is consistent with the non-zero extinction coefficient at 405~nm visible in Figure \ref{fig:carac}c. 
\pej{The calculated increase of the out-of-plane lattice constant is consistent with the measured shift }
in the Raman peak position of the A\textsubscript{1}(TO\textsubscript{3}) mode in the 100~nm BTO film, near 515~cm\textsuperscript{-1}, as a function of pump laser intensity (see Figure~\ref{fig:3}b-d). 
This mode has been assigned with the response of $a$-domains~\cite{Rubio-Marcos2015}. 
The decrease in the position of the peak from 519~cm\textsuperscript{-1} to 514~cm\textsuperscript{-1} 
\pej{corresponds} to an increase in the lattice constant perpendicular to the polarization, as evidenced by Raelijariona \textit{et al.}~\cite{Aldo2015} and \pej{confirmed by } DFT calculations reported in the Supplementary Information. 
Note that, as the surface powers during Raman measurements are much higher than in interferometric measurements (\pej{by} three orders of magnitude), the saturation observed in Figure~\ref{fig:3}c and d is not expected to happen during interferometric measurements.

As a conclusion, our results \pej{show} 
that 
giant photoinduced strains \pej{reaching 1\,\%} can be achieved in BTO films. 
The photoinduced strain in the BTO film \pej{is} driven by thermalized photo-excited carriers, in agreement with Hoang \textit{et al.}~\cite{Hoang2025}\pej{, rather than ``hot'' carriers}.  
This study opens new avenues to design more efficient lead-free photostrictive materials and devices based on the well-controlled BaTiO\textsubscript{3} material platform. 
Among others, co-doping to enhance optical absorption~\cite{Hao2022} and the thermalized carrier generation, epitaxial strains, \pej{and} domain engineering are worthy pursuits to strengthen further the photostrictive performance of barium titanate-based ferroelectrics. 

\newpage










\begin{acknowledgments}
 C. P. acknowledges support from the Air Force Office of Scientific Research through award no. FA9550-24-1-0263. S. M. acknowledges partial support from Agence Nationale de la Recherche through Grant no. ANR-24-CE08-0954.  J. H. acknowledges support from the project FerrMion of the Ministry of Education, Youth and Sports, Czech Republic, co-funded by the European Union (CZ.02.01.01/00/22\_008/0004591) 
\end{acknowledgments}

\section*{Methods}

\textit{Thin film growth \& characterization.} The BaTiO3 (BTO) and SrRuO3 (SRO) thin films have been grown with  
pulsed laser deposition using a KrF excimer laser (wavelength 248~nm) on (001)-oriented SrTiO3 substrates. Both layers were grown at 4~Hz pulse frequency, a fluence of 2~J/cm², while a temperature of 740°C  
and oxygen pressure of 0.05~mbar were used for BTO (respectively 0.2~mbar and 690°C for SRO). 
The reciprocal space maps were measured with the PANanalytical X-pert PRO MRD, the incident beam contains the Cu$_{K\alpha1}$ and Cu$_{K\alpha2}$ wavelengths. The acquisition is performed in terms of $2\theta-\omega$ scans centered on the (103) peak of the BTO film, lasting a total of eight hours and thirty-five minutes for both samples. \\

\textit{Photostriction measurements.} 
Photostriction measurements are performed with a homemade interferometer \cite{Thèse, Brevet}, using a 594nm continuous-wave diode-pumped laser from the Cobolt 04-01 series that emits up to 50mW as a probe. The interference signal is collected by a photodiode (Thorlabs PDA36A2) and is treated by a lock-in amplifier (Zurich Instruments MFLI 500kHz). The interferometer is maintained at its detection maximum using a feedback loop, relying on a microcontroller (RedPitaya STEMlab 125-14), driving a piezoelectric stack (Thorlabs PA25LEW) through a voltage amplifier (Pendulum A600). Feedback is performed with the PyRPL software \cite{pyrpl}.  The pump laser is from the Cobolt 06-01 series and emits at 405nm, with a maximum output of 120mW. Its intensity is modulated following a sine wave by an arbitrary signal generator (Tektronix AFG1022), which also serves as a reference signal for synchronous detection. The incident surface power is calculated based on the frequency component (measured via synchronous detection) of the laser output and the spot size. The sample holder has an aperture larger than the beam size, allowing the transmitted beams to exit the setup without back-reflecting in the sample and in the interferometer.

The measurements are performed at a frequency of 10~Hz, the magnitude is decreased in ten steps between the maximum and 0. The pump incidence angle with the sample surface normal is 30°. 
Due to a change in the experimental configuration, the maximum power is 33~W.cm\textsuperscript{-2} for the BTO/STO and STO samples; and 26.5~W.cm\textsuperscript{-2} for the BTO/SRO/STO and SRO/STO samples. The global uncertainty on the displacement is calculated on the basis of the uncertainty on the reference interference pattern and standard deviation during the measurement. The errors are then propagated through the classical interference formula, and the 99\% confidence interval is evaluated by taking 2.5 times the calculated error. The confidence intervals vary between 5 and 80pm. 

Since the behaviour is expected to be linear, a linear regression is performed (lines in Figure \ref{fig:3}a)). The regression gives the following coefficients for the various data sets: 
\begin{table}[h]
    \centering
    \begin{tabular}{|c|c|}
    \hline
    \textbf{Sample}  & $\mathbf{R^2}$ \\
    \hhline{|=|=|} 
    100~nm BaTiO\textsubscript{3} stack  & 0.9969 \\ 
    \hline
     50~nm SrRuO\textsubscript{3} on SrTiO\textsubscript{3} & 0.9842 \\ 
    \hline
    55~nm BaTiO\textsubscript{3} on SrTiO\textsubscript{3} & 0.9997 \\ 
    \hline
    SrTiO\textsubscript{3} & 0.9967 \\ 
    \hline
    \end{tabular}
    \caption{Estimate of the thermally-induced surface displacement}
    \label{tab:R2_photostrict}
\end{table}

\textit{Ellipsometry measurements.} 
All the films have been characterized by Variable Angle Spectroscopic Ellipsometry (VASE) in the ultraviolet to near-infrared range (UV-vis-NIR, from 0.7\,eV to 5.9\,eV) using a J. A. Woollam M2000XI. The ellipsometric angles $\Psi$ and $\Delta$ were measured at the incident angles of 50°, 60° and 70°, and the data were analyzed using the CompleteEase software supported by the J. A. Woollam Company. The ellipsometric measurements, the procedure and the optical models developed in this work to take into account the different layers are described in details in the Supplementary Information. \\

\textit{Transport measurements.} Semi-transparent Pt electrodes (8~nm thick) are deposited on the thicker 100~nm BTO film. Photocurrent and photoconductivity measurements are performed with a Keithley 2635B sourcemeter on 50~µm x 50~µm square electrodes. The light source is the same than the one used for photostriction, in static mode.  \\

\textit{Piezoelectric measurement.} The probe beam of the interferometer is focused on one of the 50~µm x 50~µm Pt electrode with the help of a x10 objective. The arbitrary signal generator (Tektronix AFG1022) is used to apply a sinusoidal voltage between the top (Pt) electrode and the bottom (SRO) electrode. The out-of-plane surface displacement is measured via interferometry. 

\textit{Temperature measurement.}  An infrared camera (FLIR A35) is used to measure the temperature elevation on the sample surface. The temperature is monitored in three points: under the laser spot, on the sample far from the laser spot and on the sample holder. This ensures the filtering of additional changes of temperature that are not due to the illumination. The laser is modulated with a 10~Hz frequency (to match the experimental conditions of photostriction measurement). The temperature under the laser spot increases, with a static and a dynamic component. Only the latter is considered for the analysis, since all of the values are taken after synchronous detection.

\textit{Raman spectroscopy.}  
Raman spectroscopy is performed with a Horiba Labram Soleil instrument at a wavelength of 405~nm. An objective x100 (NA=0.9, WD=1mm) is used to reduce the spot size to .5~µm, and the laser power is varied from 0.84~mW up to 11 mW to record the Raman spectra at different power intensities.

\textit{Density Functional Theory calculations.} We performed DFT calculations of tetragonal barium titanate using the plane-wave \textsc{Abinit} software package~\cite{Abinit2025}. We employ PAW pseudopotentials from the PseudoDojo repository~\cite{JTH}. The PBESol exchange-correlation functional~\cite{PBESol} was selected. A plane wave cut-off of 35~Ha was used alongside a $20\times 20 \times 20$ sampling of the first Brillouin zone. The electronic self-consistent field cycle was converged until the differences on the forces between two iterations was smaller than $10^{-7}$~Ha/Bohr. Structural relaxation of the ions and lattice vectors, performed using the Broyden-Fletcher-Goldfarb-Shanno algorithm, was considered converged when all forces were smaller than $5 \times 10^{-6}$~Ha/Bohr. A Fermi-Dirac smearing with a smearing temperature of 0.025~eV was applied. To mimic photo-excitation, we employed the constrained occupation number implemented in Ref.~\cite{Paillard2019} to constrain $n_{ph}$ electrons (resp. holes) in the conduction (resp. valence) bands in the unit cell. Phonon calculations at the $\Gamma$-point were also performed using Density Functional Perturbation Theory under different out-of-plane lattice constant. Throughout these calculations, the effect of the STO substrate was mimicked by clamping the first and second lattice vectors to those calculated in cubic STO (3.985~\AA).


\end{document}